\documentclass[final]{elsart}
\usepackage{graphicx}
\usepackage{amsmath}
\usepackage{times}

\newcommand{\erf}{\operatorname{erf}}
\renewcommand{\mid}{\mathop{|}}
\newcommand{\RIC}{\ifmmode\text{RIC}\else{\small RIC}\fi}

\makeatletter
\def\@overcaptionskip{0\p@}
\makeatother

\begin{document}
\begin{frontmatter}
\title{Escape rates in periodically driven Markov processes}

\author{Michael Schindler},
\author{Peter Talkner\corauthref{cor}},
\corauth[cor]{\texttt{Peter.Talkner@Physik.Uni-Augsburg.DE}\ (Peter Talkner)}
\author{Peter H\"anggi}
\address{Institut f\"ur Physik, Universit\"at Augsburg,
  D--86135 Augsburg, Germany}

\begin{abstract}%
We present an approximate analytical expression for the escape rate of
time-dependent driven stochastic processes with an absorbing boundary
such as the driven leaky integrate-and-fire model for neural spiking.
The novel approximation is based on a discrete state Markovian
modeling of the full long-time dynamics with time-dependent rates. It
is valid in a wide parameter regime beyond the restraining limits of
weak driving (linear response) and/or weak noise. The scheme is
carefully tested and yields excellent agreement with three different
numerical methods based on the Langevin
equation, the Fokker-Planck equation and an integral equation.%
\end{abstract}

\end{frontmatter}

\section{Introduction}

Although the solution of the stationary and unbounded
Ornstein-Uhlenbeck process has been found long ago, it is not yet
possible to give an analytic exact expression that includes
time-dependent driving and absorbing
boundaries~\cite{GoeRic74,HaeTalBor90}. Yet, such processes with a
linear restoring force an a periodic driving which terminate at a
prescribed threshold are widely used as models for numerous physical
effects.
Examples range from rupturing experiments on molecules~\cite{HumSza03}
where the time-dependence is introduced as linear movement of the
absorbing boundary up to totally different models like the leaky
integrate-and-fire (LIF) model for neuronal spiking
events~\cite{FouBru02,Tuckwell89,Ricciardi77,LinGarNei04,Lansky97}. The latter
is the application we primarily think of in this paper. The stochastic
variable stands for the cell soma's electric potential~$x(t)$ that is
changing due to a great many incoming signals from other neurons. It
is thus customary to employ a diffusion approximation for the
stochastic dynamics of~$x(t)$. The driven abstract LIF model assumes
the non-stationary Langevin dynamics (in dimensionless coordinates)
\begin{equation}\label{eq:langevin}
  \dot x(t) = -x(t) + A\cos(\omega t + \phi) + \sqrt{2 D}\, \xi(t)
\end{equation}
where the process starts at time~$t=0$ at~$x(0) = x_0$ and fires when
it reaches the threshold voltage~$x=a\equiv 1$. $\xi(t)$~is white
Gaussian noise. Here, a sinusoidal stimulus has been chosen for the
sake of convenience. The following analysis may easily be extended to
general periodic stimuli. The dynamics of the process~$x(t)$ is
equivalently described by a Fokker-Planck (FP) equation for the
conditional probability density function~(PDF) $\rho(x,t\mid x_0,0)$
in a time-dependent quadratic potential, $U(x,t) = \bigl(x -
A\cos(\omega t + \phi)\bigr)^2/2$, reading
\begin{equation}\label{eq:fokker-planck}
  \frac{\partial}{\partial t}\,\rho = L(t) \rho
   = \frac{\partial}{\partial x}\,\bigl(U^\prime(x,t) \rho\bigr) +
   D\frac{\partial^2}{\partial x^2}\,\rho\:,
\end{equation}
with the absorbing boundary and initial conditions
\begin{align}\label{eq:bc}
  \rho(a,t\mid x_0,0) &= 0\quad\text{for all $t$ and $x_0$}\\
  \label{eq:initial}
  \rho(x,0\mid x_0,0) &= \delta(x-x_0).
\end{align}
After firing the process immediately restarts at the instantaneous
minimum of the potential.

The set of eqs.~(\ref{eq:langevin}--\ref{eq:initial}) defines our
starting point for obtaining the firing statistics of this driven
neuron model.
Our main objective is to develop an accurate analytical approximation
that avoids certain restrictive assumptions of prior attempts. Those,
in fact, all involve the use of either of the following limiting
approximation schemes such as the limit of linear response theory
(i.e. a weak stimulus $A\ll 1$) \cite{LinGarNei04,LinSch01} or the
limit of asymptotically weak noise
\cite{GamHaeJun98,LehReiHae00a,NikStoBul03,LehReiHae00b,LehReiHae03}.
Our scheme detailed below yields novel analytic and tractable
expressions beyond the linear response and weak noise limit; as will
be demonstrated, this novel scheme indeed provides analytical formulae
that compare very favorably with precise numerical results of the full
dynamics in eqs. (\ref{eq:langevin},
\ref{eq:fokker-planck}--\ref{eq:initial}). The arguments given for the
agreement of the first-passage time distribution also hold for the
residence-time~\cite{SchTalHae04} which is not further considered here.

\section{Reduction to a discrete model}

The periodicity of the external driving with the
period~$T=2\pi/\omega$ allows one to represent the time-dependent
solution $\rho(x,t)$ of the Fokker Planck equation in terms of Floquet
eigenfunctions and eigenvalues of the Fokker-Planck operator,
$v_i(x,t)$ and $\mu_i$, respectively, \cite{GamHaeJun98,Jung93}
\begin{equation}\label{floq}
  -\frac{\partial}{\partial t} v_i(x,t) + L(t) v_i(x,t)
  = \mu_i v_i(x,t),
\end{equation}
where the eigenfunctions are periodic in time, integrable in $x$ from
$-\infty$~to $a$, and fulfill the absorbing boundary condition at
$x=a$
\begin{equation}
  v_i(a,t)=0.
\end{equation}
The time-dependent PDF can be written as a weighted sum of the Floquet
eigenfunctions
\begin{equation}
  \rho(x,t) = \sum_i c_i\, v_i(x,t)\,\exp(\mu_i t)
\end{equation}
where the coefficients $c_i$ are determined by the initial PDF. Note
that because of the absorbing boundary condition at $x=a$ the total
probability is not conserved and therefore all Floquet eigenvalues
have a non-vanishing negative real part.

The first main assumption that we impose concerns the value of the
potential at the boundary: The minimum of the potential must always
belong to the ``allowed'' region left of the threshold, and, moreover,
the potential difference between threshold and minimum, denoted by
$\Delta U(t)$, must always be larger than at least a few $D$,
i.e.~$\Delta U(t)/D > 4$. This assumption implies an exponential
time-scale separation between the average time $\tau_\kappa$ in which
the threshold is reached from the minimum of the potential compared to
the time $\tau_r$ of the deterministic relaxation towards the
potential minimum. In the dimensionless units used here $\tau_r = 1$.
For the Floquet spectrum this implies the presence of a large gap
between the first eigenvalue~$\mu_1$ which is of the same order as
$-1/\tau_\kappa$ and the higher ones which are of the order $-1$~or
smaller. After a short initial time of the order~$1$, all
contributions from higher Floquet eigenvalues can be neglected and
only the first one survives:
\begin{equation}\label{eq:mueins}
  \rho(x,t) \approx  v_1(x,t)\,\exp(\mu_1 t)
\end{equation}
In general, the Floquet eigenfunctions and the corresponding
eigenvalues are difficult to determine. A formal expansion in terms of
the instantaneous eigenfunctions $\psi_i(x,t)$ of $L(t)$ fulfilling
\begin{equation}\label{eq:psidef}
  L(t) \psi_k(x,t) = \lambda_k(t) \psi_k(x,t)
\end{equation}
is always possible though not always helpful
\begin{equation}\label{eq:upsi}
  v_i(x,t) = \sum_k d_{ik}(t) \psi_k(x,t).
\end{equation}
The periodicity of $v_i(x,t)$ and $\psi_k(x,t)$ implies that the
coefficients~$d_{ik}(t)$ also are periodic functions of time.
Expansion~(\ref{eq:upsi}), together with the Floquet
equation~(\ref{floq}), yields a coupled set of ordinary differential
equations for the coefficients~$d_{ik}(t)$ \cite{Talkner99}
\begin{equation}\label{eq:cik}
  \dot{d}_{ik}(t) - \left ( \lambda_k(t) - \mu_i \right ) d_{ik}(t)
  = \sum_l d_{il}(t) \Bigl\langle \frac{\partial}{\partial t} \varphi_k(t),
    \psi_l(t) \Bigr\rangle,
\end{equation}
where $\varphi_k(x,t)$ denotes the instantaneous eigenfunction of the
backward operator~$L^+(t)$ belonging to the eigenvalue~$\lambda_k(t)$
\begin{equation}
  L^+(t) \varphi_k(x,t) = \lambda_k(t) \varphi_k(x,t).
\end{equation}
The eigenfunctions $\psi_k(x,t)$ and $\varphi_k(x,t)$ constitute a
bi-orthogonal set of functions that always can be normalized such that
\begin{equation}
  \left\langle \varphi_l(t), \psi_k(t) \right\rangle = \delta_{kl}.
\end{equation}
Here, the scalar product $\left\langle f, g \right\rangle$ is defined
as the integral over the real axis up to the threshold:
\begin{equation}\label{sp}
  \left\langle f, g \right\rangle = \int_{-\infty}^a dx\: f(x) g(x)
\end{equation}
With our second assumption we require that the driving
frequency~$\omega$ is small compared to the relaxation rate in the
parabolic potential. Under this condition, the matrix elements
$\left\langle \partial \varphi_k(t) / \partial t, \psi_l(t)
\right\rangle$ that are proportional to the frequency~$\omega$ are
also small and may be neglected to lowest order in the equations for
the coefficients~$d_{ik}(t)$ \cite{Talkner99}. The resulting equations
are uncoupled and readily solved to yield with the periodic boundary
conditions
\begin{equation}\label{eq:deltaeinsk}
  d_{1k}(t) \approx \delta_{1k}
  \exp \Bigl( \int_0^t dt^\prime\, \lambda_1(t^\prime) - \mu_1 t\Bigr) ,
\end{equation}
where $\mu_1 = \frac{1}{T}\int_0^T \lambda_1(t)\,dt$ follows from the
periodicity of~$d_{11}(t)$.
Together with eqs.~(\ref{eq:mueins})~and (\ref{eq:upsi}) we obtain for
the long-time behavior of the~PDF
\begin{equation}
  \rho(x,t) \approx
  \exp\Bigl(\int_0^t dt^\prime\, \lambda_1(t^\prime)\Bigr) \:\psi_1(x,t).
\end{equation}
Note, that the first Floquet eigenvalue has canceled. The lowest
instantaneous eigenfunctions $\psi_1(x,t)$~and $\varphi_1(x,t)$ are
related by
\begin{equation}
  \psi_1(x,t) = \varphi_1(x,t) \rho_0(x,t),
\end{equation}
where
\begin{equation}
  \rho_0(x,t) \propto \exp\bigl(-U(x,t)/D\bigr).
\end{equation}
For the corresponding eigenvalue~$\lambda_1(t)$ we find
from~(\ref{eq:psidef})
\begin{equation}\label{eq:k1}
  \lambda_1(t) = \frac{\int_{-\infty}^a
  dx\:\varphi_1(x,t) L(t) \varphi_1(x,t)
  \rho_0(x,t)} {\int_{-\infty}^a dx\: \varphi_1^2(x,t) \rho_0(x,t)}\;.
\end{equation}
An explicit expression, valid for high potential differences, can be
given after linearization of~$U$ about~$a$
\begin{equation}
  \varphi_1(x,t) = 1 - \exp\bigl((x-a) U^\prime(a,t) / D\bigr)
\end{equation}
which gives for $\lambda_1(t)$
\begin{equation}\label{eq:rate_erf}
  \lambda_1(t) = -\frac{\Delta U(t)}{D}\; \frac{1 -
  \erf\bigl(\sqrt{\Delta U(t)/D}\,\bigr)}{1 - \exp(-\Delta U(t)/D)}\;.
\end{equation}
where $\erf(z)$ is the error function.

The waiting-time probability~\cite{TalLuc04} can be expressed as
\begin{equation}
  P(t) = \int_{-\infty}^a dx\, \varphi_1(x,t) \rho(x,t)
   = \exp\bigl(\int_0^t dt^\prime\, \lambda_1(t^\prime)\bigr)\:.
\end{equation}
Therefore, the eigenvalue~$\lambda_1(t)$ coincides with the negative
of the time-dependent escape rate~$\kappa(t)$.

With the expression~(\ref{eq:rate_erf}) for the escape rate we can
calculate the property of interest, namely the PDF for the
first-passage time~(FPT) of the attracting "integrating" state that
covers the domain $-\infty<x(t)<a$. The FPT-PDF is given by the
negative rate of change of the waiting time probability, i.e.
\begin{equation}\label{eq:fptpdf}
  g(t) = - \frac{d P(t)}{dt}
   = \kappa(t)\:\exp\Bigl(-{\int_0^t}
       \kappa(t^\prime)\,dt^\prime\Bigr)\:,
\end{equation}

The quantitative validity of these expressions for an extended
parameter regime will be checked next.

\section{Numerical analysis}%
\begin{figure}[b]%
  \centering
  \includegraphics{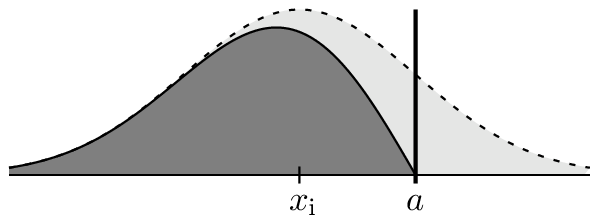}%
  \caption{The transition probabilities $p(x_{i{+}1},t+\delta t\mid
  x_i, t)$ (black line) and $N_1(x_{i{+}1},t+\delta t\mid\allowbreak
  x_i, t)$ (dashed line) from $x_i$ for a single time-step, with and
  without the absorbing boundary, respectively. The vertical line
  indicates the boundary.}%
  \label{fig:step}%
  \bigskip
  
%
  \centering
  \makebox[0.5\linewidth]{\hss\includegraphics{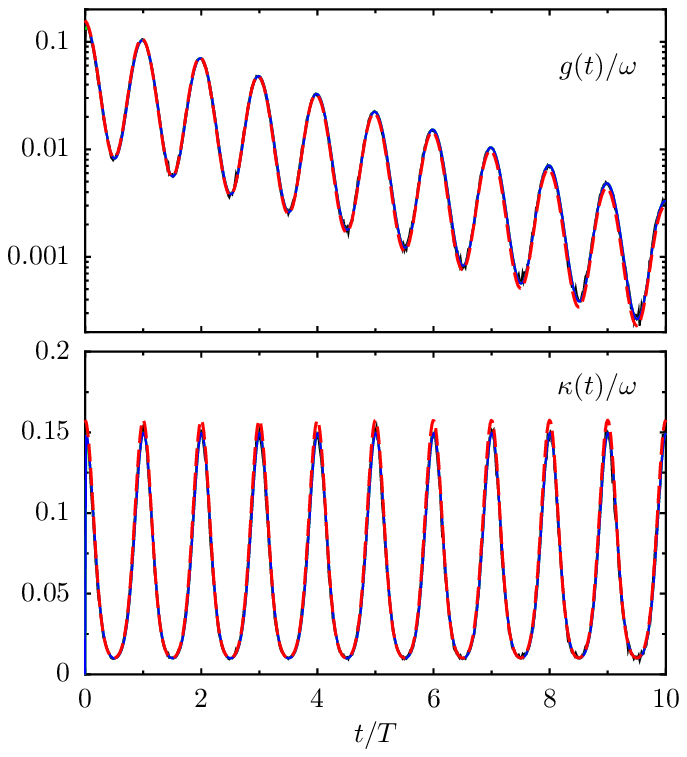}}%
  \makebox[0.5\linewidth]{ \hss\includegraphics{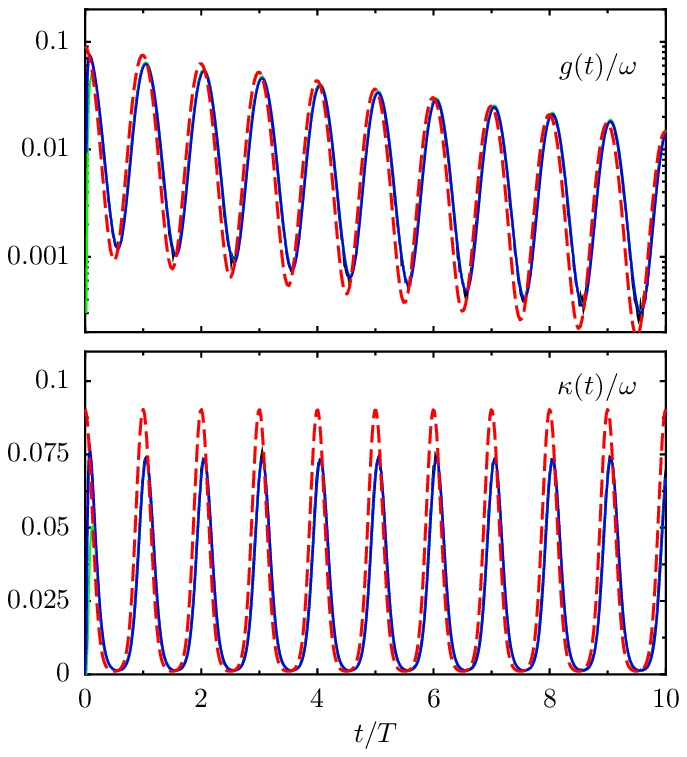}}%
  \caption{First-passage time density~$g(t)$ (upper) and
  rate~$\kappa(t)$ (lower plots) as functions of time. Displayed are
  all three numerical methods we used for testing (solid lines) and
  the approximation based on~(\ref{eq:rate_erf}) (dashed lines). The
  parameters in the left plots, $\Delta U(t)/D \in [5,8]$ and
  $\omega=0.05$, are chosen to yield a very good approximation of the
  rate by~(\ref{eq:rate_erf}). The right plots display extreme
  parameters, $\Delta U(t)/D \in [3,8]$ and $\omega=0.5$, where no
  good approximation of the rate can be expected. In both
  cases~$\phi=0$.}%
  \label{fig:fptd}%
\end{figure}%
We implemented three different numerical methods to obtain both the
FPT-PDF and the rate in order to have a reliable basis for comparison
with the analytical expression~(\ref{eq:rate_erf}).
The first method performs explicit time-steps of the Langevin
equation~(\ref{eq:langevin}). We used an elaborate technique for the
time-integration of the fluctuating force~$\xi(t)$. For points away
from the threshold~$a$ it is sufficient to take a normal distributed
random variable for the displacement due to~$\xi(t)$. Quite the
contrary in the vicinity of the absorbing boundary. Here, the integral
of~$\xi(t)$ rather behaves like a Wiener process with absorbing
boundary, as illustrated in Fig.~\ref{fig:step}. The appropriate
transition distribution, is known analytically as the weighted
difference between two normal distributions~\cite{GoeRic74}
\begin{equation}\label{eq:abswiener}
\begin{split}
  p(x_{i{+}1},t+\delta t\mid x_i, t)
  &= N_1(x_{i{+}1},\delta t\mid x_i,0)
   - N_2(x_{i{+}1},\delta t\mid x_i,0) \\
  &=: N_1(x_{i{+}1}, \delta t\mid x_i, 0)
      \big(1 - P_\text{out}(x_{i{+}1}, x_i, \delta t)\big)
\end{split}
\end{equation}
The multiplication on the right-hand side stands for a logical AND
that leads to a correction step in the algorithm. First, a new
position~$x_{i{+}1}$ is proposed according to the normal distribution
density~$N_1$. With the probability~$P_\text{out}(x_{i{+}1}, x_i,
\delta t)$ the trajectory has already crossed the boundary during this
time-step~$\delta t$ from $x_i$~to $x_{i{+}1}$ and, therefore, is to
be ended. The explicit forms of~$N_1$ and $N_2$ give
\begin{equation}\label{eq:lancorr}
  P_\text{out}(x_{i{+}1}, x_i, \delta t)
  = \frac{N_2}{N_1}
  = \exp\Bigl(-\frac{1}{D\delta t}(a-x_{i{+}1})(a-x_i)\Bigr)\:.
\end{equation}
The same formula has been given by \cite{Honerkamp90} but with a
different reasoning.

In order to get the correctly normalized FPT-PDF~$g(t)$ we counted the
number of trajectories hitting the absorbing boundary within the
interval~$[t, t+\delta t)$. The FPT-PDF is then estimated by this
number divided by~$\delta t$ and by the total number of trajectories.
The rate is given by
\begin{equation}\label{eq:numrate}
  \kappa(t) = g(t) / P(t),
\end{equation}
where $1-P(t)$ is estimated by the number of trajectories that have
escaped up to time~$t$, divided by the total number of trajectories.

For the second numerical method we have solved the FP
equation~(\ref{eq:fokker-planck}) using a Chebychev collocation method
to reduce the problem to a coupled system of ordinary differential
equations~\cite{LehReiHae00b,BerDew91}. This gives~$P(t)$ as the
integral of~$\rho(x,t)$ from $-\infty$~to $a$. The FPT-PDF in the
figures is then calculated according to eq.~(\ref{eq:fptpdf}), and the
rate again by~(\ref{eq:numrate}).

\begin{figure}[b]%
  \centering
  \makebox[0.5\linewidth]{\hss\includegraphics{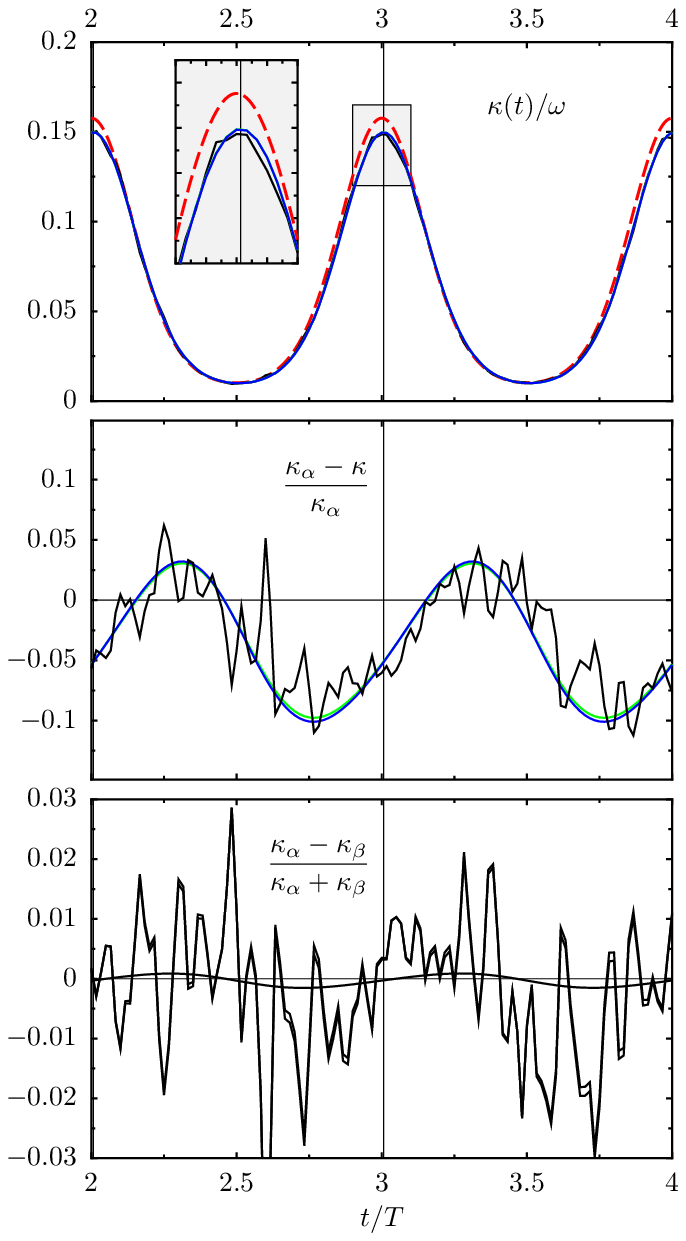}}%
  \makebox[0.5\linewidth]{\hss\includegraphics{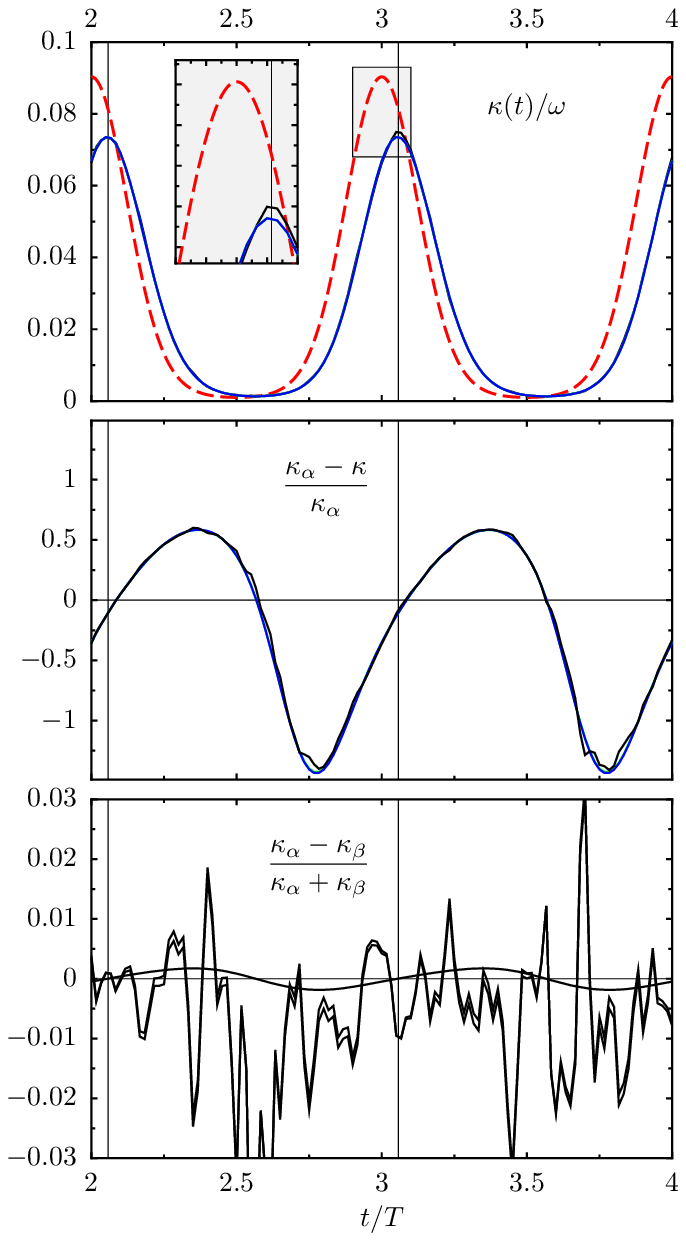}}%
  \caption{Comparison of the numerical rates and the novel
  approximation from eq.~(\ref{eq:rate_erf}). The respective
  parameters in the left/right plots are the same as in
  Fig.~\ref{fig:fptd}. Upper plots: The numerically determined rates
  are displayed as solid lines: Langevin equation simulations (black);
  Fokker-Planck equation (blue); Ricciardi's integral equation
  (green). The theoretical approximation~$\kappa(t)$ from
  eq.~(\ref{eq:rate_erf}) is displayed as the red dashed line. Middle
  plots: Relative error of the approximation $\kappa(t)$ with respect
  to each numerical rate $\kappa_\alpha(t)$ (with the same color
  coding as above). Lower plots: Errors of the numerical rates with
  respect to each other. The thin vertical lines indicate the
  positions of the numerical rates' maxima. }%
  \label{fig:highorderr}%
\end{figure}

The third method solves Ricciardi's integral equation for the FPT-PDF
and is detailed in~\cite{RicNobPir01,BuoNobRic87}. For employing his
algorithm the process must be transformed into a stationary
Ornstein-Uhlenbeck process with a moving absorbing boundary
\begin{equation}
  S(t) = a - \frac{A}{1+\omega^2}
  \Bigl[
    \cos(\omega t + \phi) + \omega\sin(\omega t + \phi) - e^{-t}
  \Bigr]\:.
\end{equation}
All three methods provide practically identical results as can be seen
in Figs.\ \ref{fig:fptd}~and \ref{fig:highorderr}. The results for the
FPT-PDF and for the rate all collapse into one single line.
Differences between the numerical methods, e.g.~fluctuations in the
histogram of the Langevin equation method are visible only in the
plots of the relative errors (Fig.~\ref{fig:highorderr} middle and
lower rows).

Figure~\ref{fig:fptd} shows that the FPT-PDF is extremely well
approximated by expression (\ref{eq:rate_erf}) for the
rate~$\kappa(t)$. In the left plots we used quite a high barrier with
quite slow driving compared to the time-scale~$\tau_r$ of the process.
Good agreement is thus expected. In the right plots we show the
situation with extreme parameters. The lower barrier height~$\Delta
U_\text{min}/D$ goes down to~$3$ where a rate-description is unlikely
to suffice. Moreover, the driving is faster, $\omega=0.5$. The system
cannot follow the driving instantaneously, and we find a shift in the
maximum of the FPT-PDF to later times. Under these conditions it is
impressing how good the novel approximation still works.

A more delicate measure for the errors of the approximation are the
rate~$\kappa(t)$ itself and its relative deviation from the three
numerically calculated rates. Both can be seen in
Fig.~\ref{fig:highorderr}. The upper row of plots shows the
approximation error of the rate for the same two parameter sets as in
Fig.~\ref{fig:fptd}. Especially at the maximum the rate is
over-estimated. This leads to a faster decay of the FPT-PDF which is
scarcely visible in Fig.~\ref{fig:fptd}. Also, the shift of the maxima
(indicated by vertical lines) can be observed. It is negligibly small
for $\omega=0.05$ but more pronounced for $\omega=0.5$.
\begin{figure}[p]%
  \centering
  \makebox[0.5\linewidth]{\hss\includegraphics{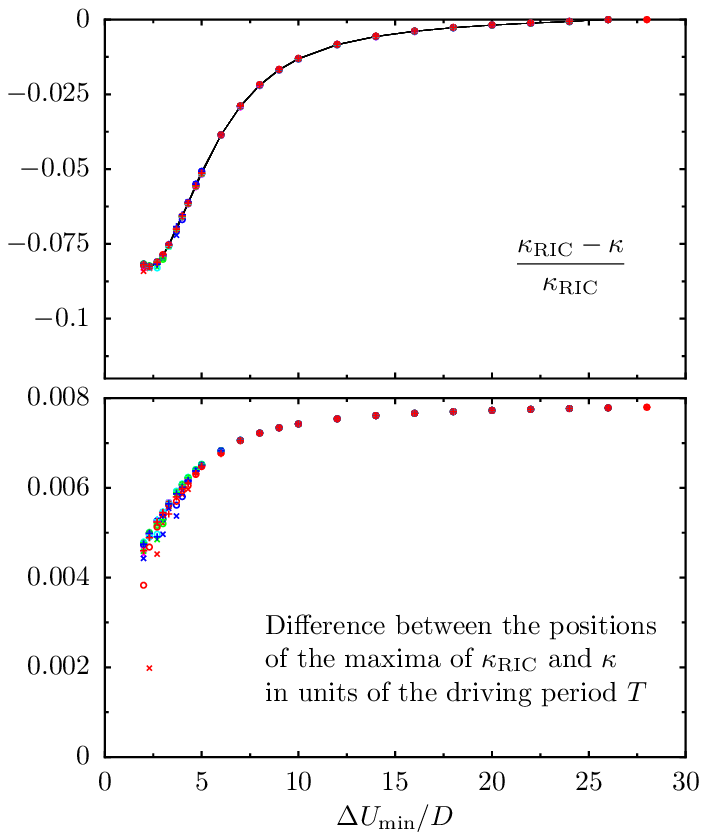}}%
  \makebox[0.5\linewidth]{\hss\includegraphics{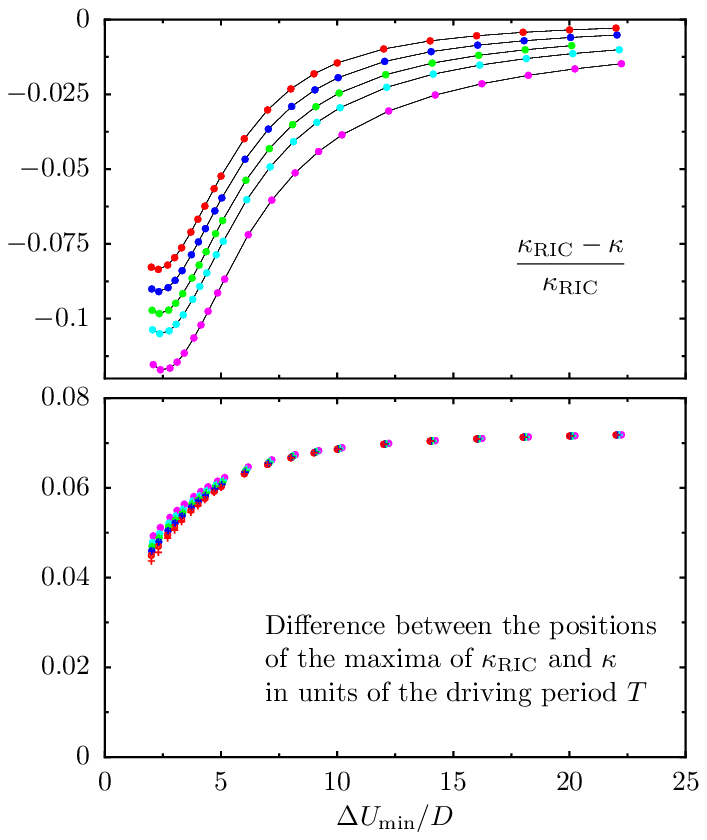}}%
  \caption{Relative error and relative time-shift of the
  rate~(\ref{eq:rate_erf}). As the basis of comparison we used the
  rate~$\kappa_\RIC$ obtained by solving Ricciardi's integral
  equation~\cite{RicNobPir01} at times $t>\tau_r$ where it has become
  periodic. Upper plots: The error relative to~$\kappa_\RIC$ evaluated
  at the maxima of~$\kappa_\RIC$.  Shown are data for $(\Delta
  U_\text{max}/D- \Delta U_\text{min}/D) \in (0.1,1,2,3,5)$, from top
  to bottom with the colors $(\text{red}, \text{green}, \text{blue},
  \text{cyan}, \text{magenta})$, and the phase $\phi\in(0, \pi/2,
  \pi)$ with the symbols $(\boldsymbol\times$, \raise -0.25ex
  \hbox{{\Large$\circ$}}, $\boldsymbol+)$. In the left panel the
  driving is slow, $\omega=0.05$, in the right it is fast,
  $\omega=0.5$. Note that the relative error is of the same order of
  magnitude for slow and for fast driving. A dependence on the phase
  $\phi$ cannot be observed. Lower plots: The difference of the
  maxima's position of~$\kappa_\RIC$ and rate~(\ref{eq:rate_erf}) in
  units of the period~$T$, again for $\omega=0.05$ (left panel) and
  $\omega=0.5$ (right panel). Color and symbol codings are the same as
  above.}%
  \label{fig:loworderr}%
  \bigskip\bigskip

  \makebox[0.5\linewidth]{\hss\includegraphics{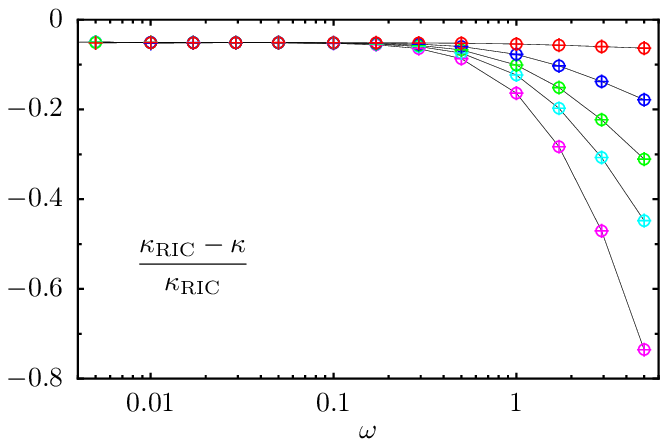}}%
  \makebox[0.5\linewidth]{\hss\includegraphics{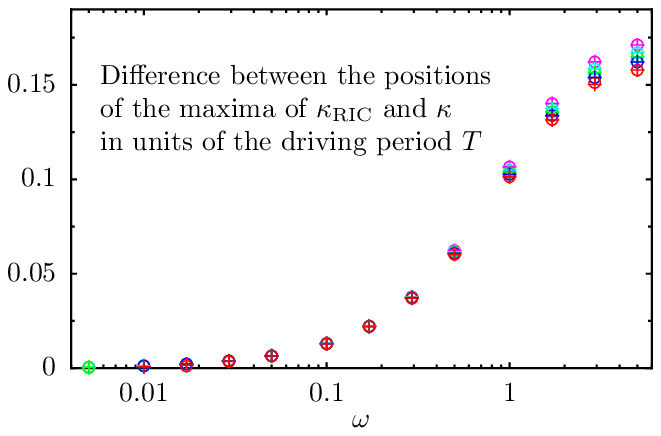}}%
  \caption{Relative error and relative time-shift of the maxima as a
  function of~$\omega$. For all data points $\Delta U_\text{min}/D =
  5$. The color and symbol codings for $\Delta U_\text{max}/D$ and
  $\phi$ are the same as in Fig.~\ref{fig:loworderr}. For fast driving
  the relative error starts depending on the interval length
  of~$\Delta U(t)/D$.}%
  \label{fig:loworderr_omega}%
\end{figure}%

In the middle row of Fig.~\ref{fig:highorderr} a systematic error of
the approximation becomes visible. The relative error with respect to
the numerical results behaves roughly sinusoidally with a phase-shift
of~$\pi/2$ relative to the driving and with an additional constant
offset. For the instantaneous rate expression~(\ref{eq:k1}) to be
valid it is necessary that the driving signal is sufficiently slow. If
this assumption is violated then a rate can still be defined if the
barrier is sufficiently high. But in addition to the leading
term~$d_{11}(t)$ in~(\ref{eq:upsi}) the higher instantaneous
eigenfunctions must be taken into account. The coupling to the
coefficients~$d_{1k}(t)$ is induced by the matrix
elements~$\langle\dot\phi_k(t), \psi_l(t)\rangle$, see
eq.~(\ref{eq:cik}), containing a time derivative that introduces
non-adiabatic corrections to the rate and, consequently, to the
statistics of the~FPT.

It is quite astonishing, that the huge relative error in the right
middle plot of Fig.~\ref{fig:highorderr} leads to such a good result
in Fig.~\ref{fig:fptd}. The explanation for this is that the
FPT-PDF~(\ref{eq:fptpdf}) uses the time-integrated rate. Therefore,
errors are important only where the rate is large. A closer look on
the plot shows that around the maxima of the rate the relative error
is comparably small. Because the errors are linear in time around the
rate's maxima they cancel out when integrated over time
in~(\ref{eq:fptpdf}). The same is valid for the residence time whose
PDF also contains integrals of the rate~\cite{SchTalHae04,Talkner03}.

Figure~\ref{fig:loworderr} shows this relative error of~$\kappa(t)$ at
the maxima of the numerically obtained rate as a function of the
barrier height. Again, two different driving frequencies are given. In
both cases the relative error has the same order of magnitude, and
thus explains why both parameter sets in Fig.~\ref{fig:fptd} yield
good approximations.

Finally, we would like to point the reader's attention to the
limitations of the linear response approximation. For linear response
the parameter ratio~$A/D$ needs to be small. In our validating example
in Fig.~\ref{fig:fptd} (left plots) it takes on the value $A/D=1.5$.
Thus, our approximation scheme is valid beyond the linear response
limit.

The time-scale of the driving force is mainly restricted by the
relaxation time-scale~$\tau_r$ and much less by the magnitude of the
rate itself. There is no restriction on the relative magnitudes of
$\kappa$~and $T$. Instead, both $\omega$ and $\kappa$ have to be
sufficiently small. Fig.~\ref{fig:loworderr_omega} indicates that both
the relative error and the time-shift of the maxima's positions are
modest for~$\omega<0.1$.

\section{Conclusions}

By reference to a \textit{discrete} Markovian dynamics for the
corresponding full space-continuous stochastic process we succeeded in
obtaining an analytical approximation for the time-dependent escape
rate which can be used for calculating first-passage time statistics.
This result is valid beyond the restraining limits of linear response
or asymptotically weak noise and of adiabatically slow driving.

We checked our findings using simulations of the Langevin
equation~(\ref{eq:langevin}) and numerical solutions of the equivalent
FP equation in~(\ref{eq:fokker-planck}) and of the integral equation
in~\cite{RicNobPir01}. We found an impressive agreement for the
first-passage time density and a good match for the rate which is the
more delicate property for comparison.

Finally, we note that our method is not restricted to a periodic
forcing but applies also to arbitrary drive functions. However, in the
oscillatory case some of the approximation errors cancel out. This
leads to useful results in extreme parameter regimes where agreement
cannot be expected \textit{a priori}.

This work has been supported by the Deutsche Forschungsgemeinschaft
via project HA1517/13-4 and SFB-486, projects A10 and B13.

\end{document}